\documentclass[prl,twocolumn,superscriptaddress,showpacs,amsmath,amstex,amssymb,citeautoscript]{revtex4-1}
\pdfoutput=1
\usepackage{natbib}
\usepackage[english]{babel}
\usepackage{letltxmacro}
\usepackage{latexsym}
\LetLtxMacro{\ORIGselectlanguage}{\selectlanguage}
\makeatletter
\DeclareRobustCommand{\selectlanguage}[1]{%
  \@ifundefined{alias@\string#1}
    {\ORIGselectlanguage{#1}}
    {\begingroup\edef\x{\endgroup
       \noexpand\ORIGselectlanguage{\@nameuse{alias@#1}}}\x}%
}
\newcommand{\definelanguagealias}[2]{%
  \@namedef{alias@#1}{#2}%
}
\makeatother
\definelanguagealias{en}{english}
\definelanguagealias{English}{english}
\usepackage{graphicx}
\usepackage{amsmath}
\usepackage{amsfonts}
\usepackage{amssymb}
\usepackage{bm}
\usepackage{color}
\usepackage[percent]{overpic}
\usepackage{soul} 
\usepackage{amssymb}
\usepackage{wasysym}
\usepackage{dsfont}
\usepackage{float}
\usepackage[mathscr]{euscript}

\usepackage{hyperref}
\hypersetup{
    pdfstartview={FitH},    % fits the width of the page to the window
    colorlinks=true,       % false: boxed links; true: colored links
    linkcolor=blue,          % color of internal links (change box color with linkbordercolor)
    citecolor=blue,        % color of links to bibliography
    filecolor=magenta,      % color of file links
    urlcolor=blue           % color of external links
}

\newcommand{\be}{\begin{equation}}
\newcommand{\ee}{\end{equation}}
\newcommand{\bea}{\begin{eqnarray}}
\newcommand{\eea}{\end{eqnarray}}

\newcommand{\mc}{\mathcal}

\newcommand{\bra}[1]{\langle {#1} |}
\newcommand{\ket}[1]{| {#1} \rangle}
\newcommand{\vect}[1]{\boldsymbol{#1}}

\usepackage{graphicx}
\usepackage[colorinlistoftodos]{todonotes}
\usepackage{verbatim}
\usepackage[normalem]{ulem}

\usepackage{pdfpages}
\usepackage{pgffor}
\makeatletter
\AtBeginDocument{\let\LS@rot\@undefined}
\makeatother

\begin{document}

\title{Complex density wave orders and quantum phase transitions in a model of square-lattice Rydberg atom arrays}
\author{Rhine Samajdar}
\affiliation{Department of Physics, Harvard University, Cambridge, MA 02138, USA}
\author{Wen Wei Ho}
\affiliation{Department of Physics, Harvard University, Cambridge, MA 02138, USA}
\author{Hannes Pichler}
\affiliation{Department of Physics, Harvard University, Cambridge, MA 02138, USA}
\affiliation{ITAMP, Harvard-Smithsonian Center for Astrophysics, Cambridge, MA 02138, USA}
\affiliation{Division of Physics, Mathematics and Astronomy, California Institute of Technology, Pasadena, CA 91125, USA}
\author{Mikhail D. Lukin}
\affiliation{Department of Physics, Harvard University, Cambridge, MA 02138, USA}
\author{Subir Sachdev}
\affiliation{Department of Physics, Harvard University, Cambridge, MA 02138, USA}

\begin{abstract}
We describe the zero-temperature phase diagram of a model of a two-dimensional square-lattice array of neutral atoms, excited into Rydberg states and 
interacting via strong van der Waals interactions. Using the density-matrix renormalization group algorithm, we map out the phase diagram and obtain a rich variety of phases featuring complex density wave orderings, upon varying lattice spacing and laser detuning.
While some of these phases result from the classical optimization of the van der Waals energy, 
we also find intrinsically quantum-ordered phases stabilized by quantum fluctuations. These phases are surrounded 
by novel quantum phase transitions, 
which we analyze by finite-size scaling numerics and Landau theories.
Our work highlights Rydberg quantum simulators in higher dimensions as promising platforms to realize exotic many-body phenomena. 

\end{abstract}
%\date{\today}
\maketitle

\hypersetup{linkcolor=blue}

%\emph{Introduction.}---
The ability to fully control coherent quantum many-body systems is an exciting frontier.
Apart from quantum information processing, controlled many-body systems can enable 
new insights into 
strongly correlated phases of matter including the realization of exotic orders, nonequilibrium quantum dynamics, and
the role of quantum entanglement. A number of physical platforms, such as cold atoms~\cite{bloch2008many,bernien2017probing}, trapped ions~\cite{bruzewicz2019trapped}, and superconducting qubits~\cite{wendin2017quantum}, have exhibited these capabilities
in  systems of small to intermediate sizes. In this regard, arrays of neutral atoms trapped in optical tweezers and interacting via controlled excitations into atomic Rydberg states, provide an especially promising platform. % for these studies.
A unique feature of this system is the ability to arrange atoms in arbitrary geometries in one~\cite{endres2016atom}, two~\cite{barredo2016atom,lee2017defect, brown2019gray, de2019defect}, or three~\cite{barredo2018synthetic,kumar2018sorting} spatial dimensions. Additionally, strong (potentially direction-dependent) interactions lead to the Rydberg blockade mechanism \cite{jaksch2000fast}, preventing two nearby atoms to be simultaneously excited to the Rydberg state.
Together, these properties allow for the programmable realization and high-fidelity manipulation of a wide range of effective interacting spin models \cite{labuhn2016tunable, de2019observation}.
Indeed, experiments on one-dimensional Rydberg atom arrays have already  
shed light on various phenomena, such as the nature of quantum phase transitions (QPTs) \cite{samajdar2018numerical, whitsitt2018quantum} and the Kibble-Zurek mechanism \cite{keesling2019quantum}, and even uncovered surprising dynamical behavior such as quantum many-body scarring \cite{turner2018weak, ho2019periodic}.
The interplay of constraints from the Rydberg blockade and the geometric positions of the atoms in two spatial dimensions implies that a host of richer, nontrivial phenomena can be realized, ranging from phases with competing crystalline orders \cite{laycock2011creation, schauss2012observation, qian2013quantum, petrosyan2013two}, to the physics of quantum dimer models and topological phases \cite{de2019observation, celi2019emerging}.

\begin{figure}[t]
\includegraphics[width=0.9\linewidth, trim={0.15cm 0.85cm 0 0 },clip]{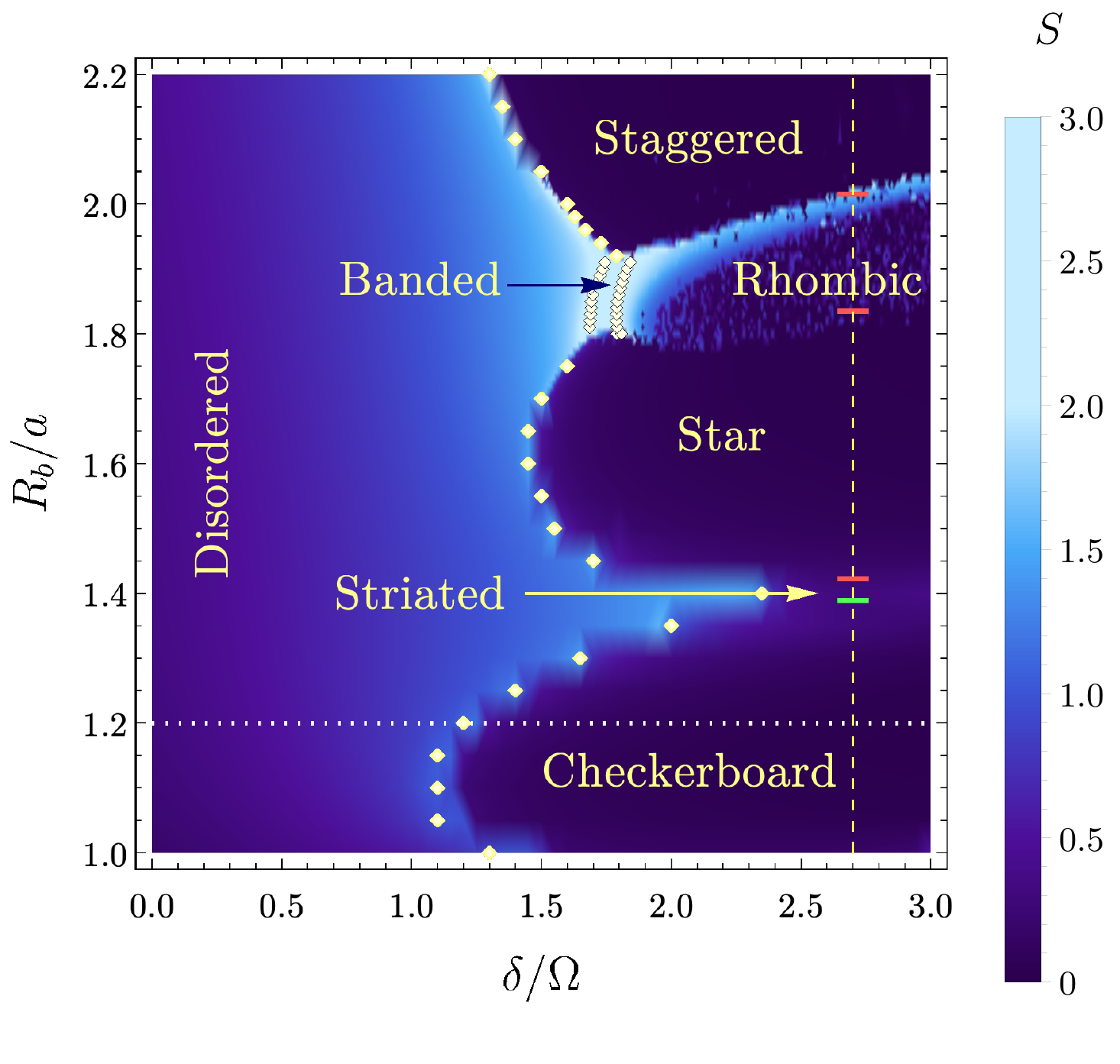}
\caption{\label{fig:PD}Phase diagram of the 2D Rydberg Hamiltonian \eqref{eq:Rydberg}, traced out by the bipartite entanglement entropy $S$ on a $15$\,$\times$\,$8$ square lattice on a cylinder. 
The five density-wave-ordered phases along the line $\delta/\Omega$\,$=$\,$2.7$ are sketched in Fig.~\ref{fig:Phases}, together with the banded phase. The QPT along the line $R_b /a$\,$=$\,$1.2$ is analyzed in Fig.~\ref{fig:Collapse}. Red (green) dashes mark first-order (continuous) transitions. The yellow diamonds demarcating the phase boundaries are the calculated finite-size pseudocritical points \cite{hamer2000finite}.
}
\end{figure}

Motivated by such possibilities, %in this Letter, 
we numerically study the phases in a fully coherent collection of Rydberg atoms on a square lattice, using the density-matrix renormalization group (DMRG) algorithm. Along with several density-wave-ordered 
phases, arising from the densest (constrained) packing of Rydberg excitations,
we also find signatures of % two 
intrinsically quantum phases 
%(the striated and banded) 
stabilized by quantum fluctuations [Fig.~\ref{fig:PD}].  
We quantitatively determine the % different 
phase boundaries and % in the process, 
map out the full phase diagram in detail.
Additionally, we comprehensively examine the nature of the QPTs, based on finite-size scaling analyses and effective Landau theories. In particular, we demonstrate an instance of a QPT in the 3D Ising universality class; we thus propose the 2D Rydberg atom array as the first experimental platform to unambiguously observe this transition that has proved elusive in condensed-matter systems to date \cite{coldea2014}. Furthermore, we provide evidence for exotic QPTs, such as those described by theories of three-dimensional O($\mathscr{N}$)-symmetric vector models with anisotropic perturbations.

 \emph{Model.}---We study the following Hamiltonian describing interacting Rydberg atoms arranged in a 2D square lattice of size $N$\,$\equiv$\,$L_x$\,$\times$\,$L_y$, with open (periodic) boundary conditions in the $x$\,$(y)$-direction:
\begin{alignat}{1}
\label{eq:Rydberg}
\nonumber H_{\rm Ryd}&=\sum_{i=1}^N \frac{\Omega}{2} \left(\ket{g}_i\!\bra{r}+\ket{r}_i\!\bra{g} \right)-\delta \ket{r}_i\!\bra{r} \\
&+ \frac{1}{2}\sum_{i\ne j} V\left(\lvert \lvert \vect{x}^{}_i-\vect{x}^{}_j \rvert  \rvert/a\right)\ket{r}_i\!\bra{r}\otimes \ket{r}_j\!\bra{r}.
\end{alignat}
Here, $i$ labels sites at positions $\vect{x}_{i}$ of the lattice (with lattice constant $a$), while $\ket{g}_i$ and $\ket{r}_i$ denote the internal atomic ground state and a Rydberg state of the $i$-th atom, respectively. The parameters $\Omega$ (Rabi frequency) and $\delta$ (detuning) characterize a coherent laser driving field, while $V(x)$\,$=$\,$C_6/x^6$ quantifies the van der Waals interactions of atoms in Rydberg states. 
$H_{\rm Ryd}$ can equivalently be parametrized by the Rydberg blockade radius, $R_b$, defined by $V (R_b/a)$\,$\equiv$\,$\Omega$, within which interactions are so strong that effectively no two neighboring atoms can simultaneously be excited to Rydberg states; the sites are then said to be blockaded \cite{jaksch2000fast}. 
Note that $H_{\rm Ryd}$ maps to a model of \textit{hard-core} bosons \cite{sachdev2002mott,fendley2004competing}. However, it does not have a global U(1) symmetry, distinguishing it from related models \cite{hebert2001quantum, yamamoto2012quantum, xiao2015hard} on the square lattice that have attracted much attention in the context of supersolidity.

The ground states hosted by the Hamiltonian $H_{\rm Ryd}$ depend sensitively on % the precise values of 
$\delta/\Omega$ and $R_b/a$, which control the density of Rydberg excitations $\langle n_i \rangle$;\,$n_i$\,$=$\,$\ket{r}_i\bra{r}$. At large negative % values of 
$\delta/\Omega$, configurations with most atoms in $\ket{g}$ are favored, resulting in a so-called \textit{disordered} phase. For large positive values of $\delta/\Omega$, the density of atoms in $\ket{r}$ increases, but their geometric arrangement is constrained by the interactions between proximate Rydberg atoms. This competition between $\delta$ and $V$ (or $R_b$) leads to ordered phases with different spatial symmetries, referred to as ``Rydberg crystals'' \cite{fendley2004competing, pohl2010dynamical} in which Rydberg atoms are arranged regularly across the array. In two spatial dimensions, classical combinatorics suggest that a plethora of such crystalline phases can be realized, in close correspondence with the solutions of the circle packing optimization problem \cite{lodi2002two}, which is known to be NP (nondeterministic polynomial time) hard \cite{demaine2016circle, pichler2018quantum, pichler2018computational}.

\emph{Methods and observables.}---We numerically obtain the ground states of $H_\text{Ryd}$ for various values of $R_b/a$,\,$\delta/\Omega$ using DMRG \cite{white1992density, white1993density} with a snakelike matrix product state ansatz
[see the Supplemental Material\,(SM) \cite{sm} for details]. 
We retain interactions between atoms separated by up to two lattice units (third-nearest neighbors) in Eq.~\eqref{eq:Rydberg}: with this truncation, one approximates the physics of $H_{\rm Ryd}$ faithfully for $R_b/a$\,$\lesssim$\,$\sqrt{5}$.
The linear dimensions $L_x$,\,$L_y$ are chosen to be compatible with most of the possible ordering patterns while respecting the optimal aspect ratio $\alpha^*$\,$=$\,$L_x/ L_y$\,$\simeq$\,$1.9$ needed to minimize finite-size corrections in $1/L_y$
and render the bulk of the cylinder a good approximation of the infinite 2D system
\cite{white2007neel}.  Unless specified otherwise, we choose $L_x$\,$=$\,$15$, $L_y$\,$=$\,$8$, and work in units of $\Omega$\,$=$\,$1$,\,$a=1$. 

\begin{figure}[b]
\includegraphics[width=1\columnwidth,trim={0 0 0 0.1cm},clip]{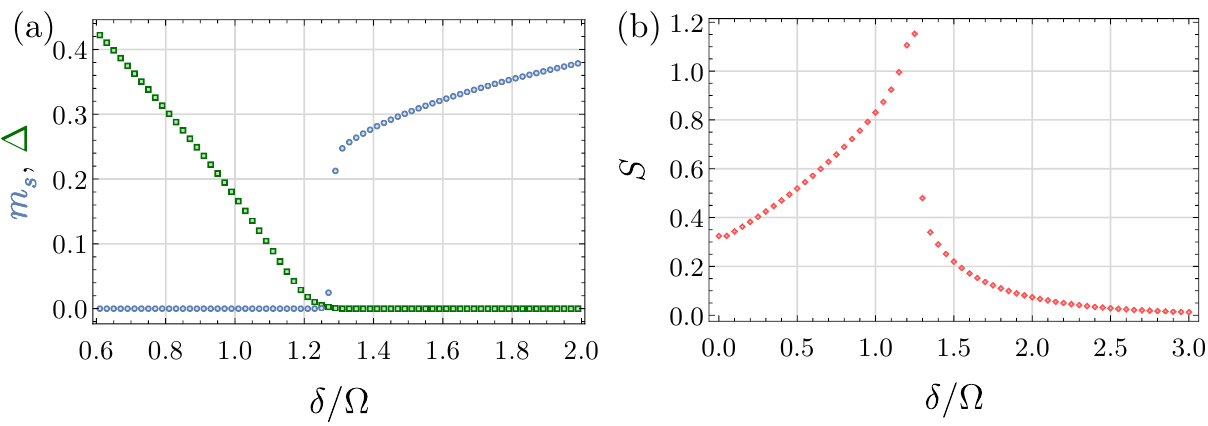}
\caption{\label{fig:Obs}The observables used to diagnose the $\mathbb{Z}_2$-symmetry-breaking QPT are (a) the staggered magnetization, the energy gap, and (b) the bipartite entanglement entropy, plotted here along the line $R_b/a = 1.2$. 
}
\end{figure}

\begin{figure*}[htb]
\includegraphics[width=1.0\linewidth, trim={0.1cm 0 0 0},clip ]{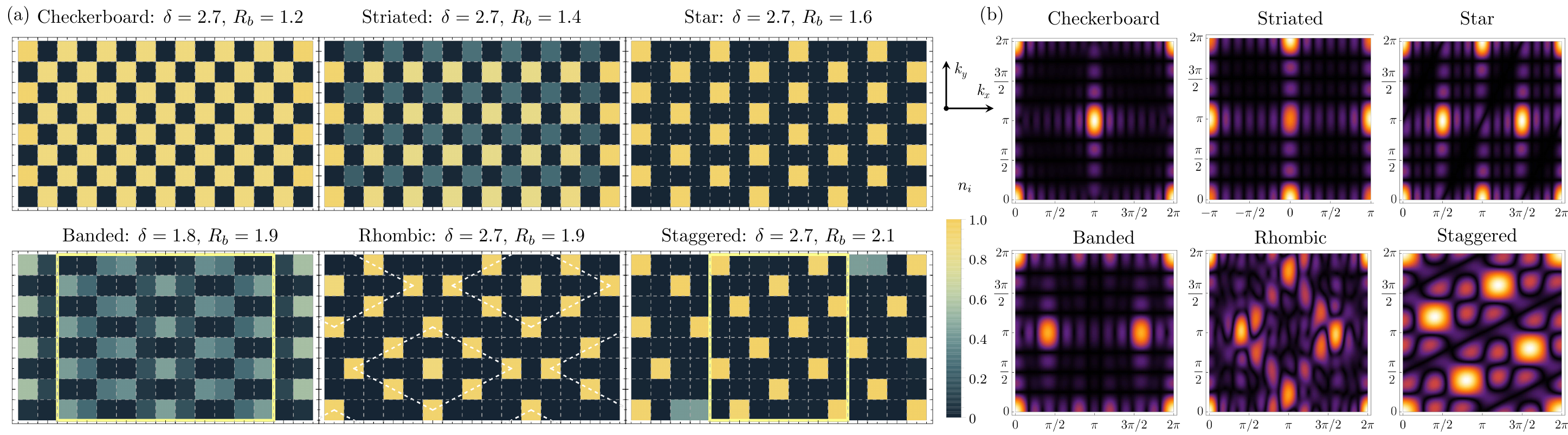}
  \caption{\label{fig:Phases}Magnetization profiles, $n^{}_i$ and $\lvert n(\vect{k})\rvert$, of the six ordered phases in (a) real and (b) momentum space. In the cases where boundary effects induce defects near the edges, the bulk (framed) reflects the ideal ordering.}
\end{figure*}

The properties used to identify the phases and the QPTs between them (Fig.~\ref{fig:Obs}) are best illustrated in a context that can be understood analytically. 
To begin, we scan $\delta$ along the line $R_b$\,$=$\,$1.2$, where only nearest-neighbor sites are blockaded. For small $\delta$, the system is a ``paramagnet'' with a unique, featureless ground state containing a low density of Rydberg excitations, whereas for larger positive $\delta$, the ground state is twofold degenerate, with an antiferromagnetically ordered, checkerboard arrangement of excitations (\textit{i.e.}, a N\'{e}el state) \cite{ji2011two, lienhard2018observing}. 
Therefore, the staggered magnetization \cite{matsubara1956lattice, Batyev1984} $m^{}_s$\,$=$\,$\langle \lvert M_N^{}\rvert\rangle;$\,$M_N^{}$\,$\equiv$\,$\sum_{i=1}^{N}(-1)^i\,(n^{}_i-1/2)/N$, can serve as an order parameter detecting the $\mathbb{Z}_2$-symmetry-breaking QPT. As $\delta$ approaches the quantum critical point (QCP) $\delta_c$ from above, $m_s$ vanishes. Moreover, approaching $\delta_c$ from below, the energy gap to the first-excited state $\Delta$\,$\equiv$\,$E_1$\,$-$\,$E_0$ closes at the QPT \cite{sachdev2011quantum, dutta2015quantum}.
This behavior is indeed corroborated by our numerics in Fig.~\ref{fig:Obs}, where the QPT occurs at $\delta_c$\,$\approx$\,$1.3$. In Fig.~\ref{fig:Obs}(a), $\Delta$ is seen to be nearly zero in the ordered phase, which is a numerical indicator of the ground state being degenerate. 

The drawback of using an order parameter such as $m^{}_s$ or the gap $\Delta$ to determine the (finite-size) phase boundaries is that the former requires \textit{a priori} knowledge of the ordering of the phases and obtaining the latter is computationally expensive.
A more unbiased diagnostic is the half-cylinder bipartite entanglement entropy (EE) of the ground state $S$\,$\equiv$\,$-\mathrm{Tr}\,(\rho_r\ln\rho_r)$, where $\rho_r$ is the  reduced density matrix of half the cylinder (partitioned along $\hat{x}$). 
As seen in Fig.~\ref{fig:Obs}, $S$ peaks near the QCP \cite{stoudenmire2015assembling} and then drops sharply in the ordered phase 
(DMRG selects one of the two symmetry-broken states rather than their superposition, being biased toward states having low entanglement \cite{stoudenmire2012studying}). 
This is indeed the quantity we scrutinize to limn the phases below, and 
generate the phase diagram. % of Fig.~\ref{fig:PD}.

 \emph{Ordered phases at larger $R_b$.}---Away from the simple case above, we find a number of new phases, with different density-wave orderings that have no analog in 1D. Figure~\ref{fig:PD} displays three broad lobes in the $(\delta/\Omega,$\,$R_b/a)$ parameter space, which, along with the intermediate regions between them, constitute six ordered phases. Their magnetization profiles as well as their associated Fourier transforms $n (\vect{k})$\,$=$\,$\sum_i \exp(i \vect{k}\cdot \vect{r}_i)$\,$\langle n_i\rangle /\sqrt{N}$ 
 % (which is a convenient metric to diagnose the ordered phases) 
 are presented in Fig.~\ref{fig:Phases}. % for each of these phases.   

%To orient the reader, 
Consider increasing $R_b$ from the checkerboard phase at $R_b$\,$\approx$\,$1.2$ while remaining on the line $\delta$\,$=$\,$2.7$ (vertical dashed line in Fig.~\ref{fig:PD}).
As $R_b$ is increased, the N\'{e}el order begins to melt, 
and the system transitions into a \textit{striated} phase \cite{batrouni2000phase, chen2008supersolidity} characterized by a nonzero row magnetization
$
m^{}_r$\,$\equiv$\,$\big\lvert\sum_{i=1}^{N}(-1)^{\mathrm{row}(i)}\left\langle n_i\right\rangle\big\rvert/N.
$
Given that in the classical limit ($\delta/\Omega$\,$\rightarrow$\,$\infty$,\,$R_b/a$\,$\ne$\,$0$), 
%where $n_i \in \{0, 1\}$,
the star state (described below) is always energetically favored over one with pure striated ordering, the appearance of this phase is unexpected. A key role is played here % in this regard 
by quantum fluctuations, %when $\delta/\Omega$\,$<$\,$\infty$,
which stabilize the striated phase in a narrow window: the system optimizes the packing fraction by placing Rydberg atoms on one sublattice in the, say, odd rows together with a small but nonzero density of delocalized excitations on the same sublattice on the even rows. Smearing out these additional excitations offsets the energy penalty due to $V$ while maximizing the reduction in energy from $\delta$. Hence, the striated ordering coexists with a vestigial N\'{e}el order.

Once $R_b$\,$\gtrsim$\,$\sqrt{2}$, diagonally adjacent sites are also blockaded by the repulsive interactions and the system enters a so-called \textit{star} phase \cite{dang2008vacancy, dang2010phases};
the order parameter is the conventional magnetization $m$\,$\equiv$\,$\sum_{i=1}^{N}\left\langle n^{}_i -1/2\right\rangle /N$.
Next to the star phase, lies the \textit{rhombic} phase in which Rydberg excitations are clustered in a pattern resembling a diamond. 
Despite the large size of the 
unit cell, consisting of 40 sites and 9 Rydberg atoms, the rhomboidal crystal is remarkably robust on a wide range of lattice sizes. 
This phase is separated from the disordered one by a sliver of the intermediate (purely quantum) \textit{banded} phase. 
Finally, increasing $R_b$ even further, till third-nearest neighbors are blockaded, brings us to the \textit{staggered} phase where nearest excitations are always a distance of $\sqrt{5}$ apart, their arrangement being reminiscent of the allowed moves of a knight on a chessboard. 
The salient features of these phases are enlisted in Table~\ref{Table:z}.

\bgroup
\def\arraystretch{1.5}
\begin{table}[t]
%\vspace*{0.4cm}
\centering
{
\bgroup
\setlength{\tabcolsep}{1pt}
\begin{ruledtabular}
\begin{tabular}{l l l l l } 
\multicolumn{1}{l}{Phase} &\multicolumn{1}{l}{{\normalsize $\hat{a}^{}_{1,2}$}} &\multicolumn{1}{l}{$\mathcal{D}$}  &\multicolumn{1}{l}{Maxima of $\lvert n(\vect{k}) \rvert$} &\multicolumn{1}{l}{{\normalsize $\bar{n}^{}_b$}} \\
\hline
Checkerboard & $\hat{x} \pm \hat{y}$ & 2  & $(\pi, \pi)$ & $\frac{1}{2}$ \\
Striated & $2 \hat{x}$, $2 \hat{y}$ & 4  & $(\pi, 0),\,(\pi, \pi)$ & -- \\
Star  & $2 \hat{x}\, \pm\, \hat{y}$ & 8& $\pm(\pi/2, \pi),\, (\pi, 0)$ & $\frac{1}{4}$ \\
Banded  & $5 \hat{x},\, \hat{2y}$& 20 & $\pm(2 \pi/5, \pi)$ & --  \\
Rhombic  & $5 \hat{x} \pm 4 \hat{y}$& 80 & $\pm (\pi,  \pi/4), \, \pm(2 \pi/5, \pi)$ & $\frac{9}{40}$  \\
Staggered  & $2 \hat{x} + \hat{y},  \hat{x} - 2 \hat{y}$ & 10 & $\pm (\frac{2 \pi}{5}, -\frac{4 \pi}{5}),\, \pm(\frac{4 \pi}{5}, \frac{2\pi}{5})$  &$  \frac{1}{5}$ \\
\end{tabular}
\end{ruledtabular}
}
\egroup
\caption{\label{Table:z}Properties of the six ordered phases for infinite system sizes: the primitive lattice vectors of the density wave $\hat{a}^{}_{1,2}$, the ground-state degeneracy $\mathcal{D}$, the dominant peaks in the Fourier spectrum, and the density of Rydberg excitations in the classical limit, $\bar{n}_b$. On an infinite lattice, the ground state also includes Rydberg crystals with $C_4$-rotated copies of the lattice vectors and momentum peaks tabulated above, even though the symmetry between rows and columns is broken by the boundary conditions in a finite-size system. }
\end{table}
\egroup

 \emph{Nature of phase transitions.}---While the array of ordered phases of the 2D Rydberg Hamiltonian \eqref{eq:Rydberg} are intriguing, % on its own, 
 equally interesting are the symmetry-breaking QPTs that engender them. % in the first place. 
 We focus on the continuous transitions---encountered upon going from the disordered phase to one of the ordered phases---that can be characterized by 
 universal critical exponents \cite{sachdev2011quantum, dutta2015quantum}, which we numerically determine by finite-size scaling (FSS) \cite{fisher1972scaling, hamer1980finite}.

\begin{figure}[b]
\includegraphics[width=1\columnwidth]{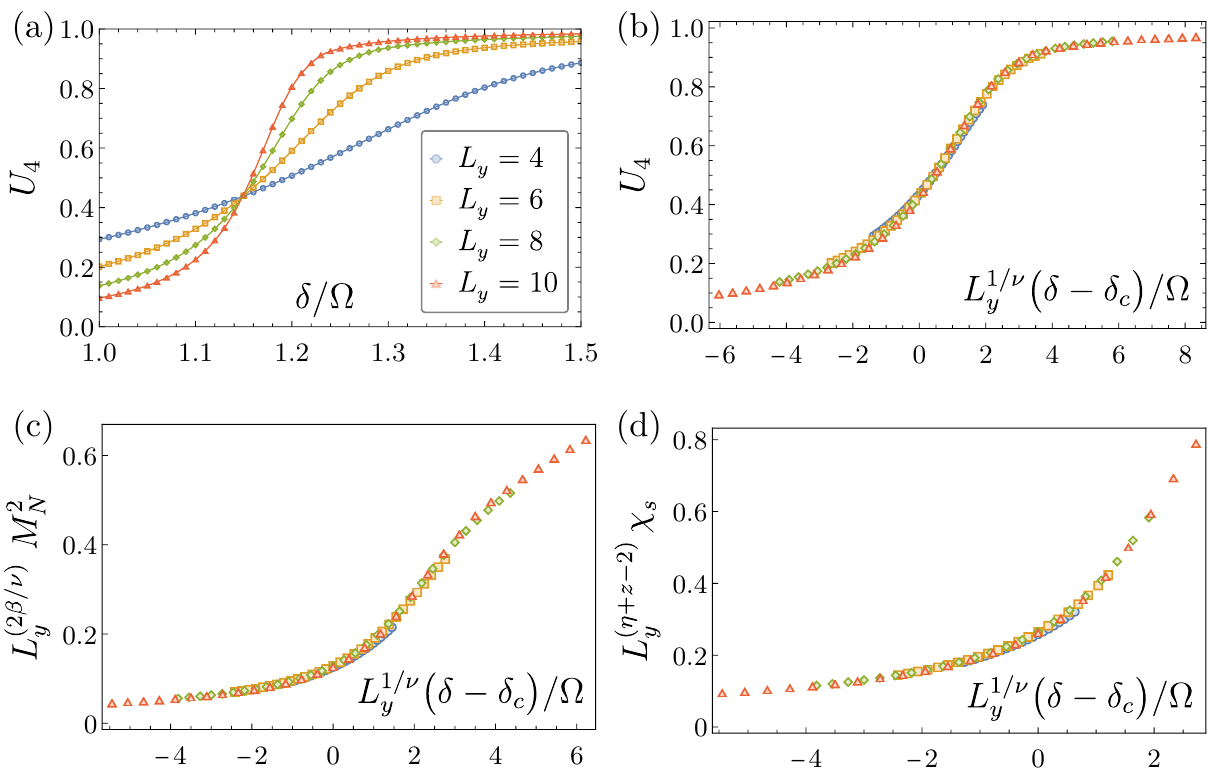}
  \caption{\label{fig:Collapse}(a) Binder cumulant $U^{}_4$ for different lattice sizes; all of the curves intersect at the QCP. Data collapse is obtained for (b) $U^{}_4$, (c) the squared magnetization, and (d) the susceptibility using the 3D Ising critical exponents in the FSS ans\"{a}tze. }
\end{figure}

We begin by examining the purportedly simplest QPT  from the disordered to the checkerboard phase on the line $R_b$\,$=$\,$1.2$ (horizontal dotted line in Fig.\,\ref{fig:PD}). First, we precisely establish the location of the QCP in the thermodynamic limit for use in all scaling forms, by computing 
% . To this end, we  compute 
the Binder cumulant \cite{binder1981finite} $2\, U^{}_4$\,$\equiv$\,$3$\,$-$\,$\langle M_N^4\rangle/\langle M_N^2\rangle^2$,
which is size independent at the QCP for sufficiently large systems. As is visible in Fig.~\ref{fig:Collapse}(a), the cumulants all intersect at one point for system sizes ranging from $L_y$\,$=$\,$4$ to $10$, with fixed aspect ratio $\alpha$\,$=$\,$2$. Crossings of the curves for pairs of system sizes $L_y$ and $L_y$\,$+$\,$2$ proffer a sequence of finite-size estimates $\delta_c\,(L)$ of the critical point, which can be extrapolated to $L_y$\,$\rightarrow$\,$\infty$ \cite{qin2015multiplicative}, yielding $\delta_c/\Omega$\,$=$\,$1.1477$\,$\pm$\,$0.0006$.

Near the QCP, the correlation length diverges as $\xi$\,$\sim$\,$\lvert\delta$\,$-$\,$\delta_c\rvert^{-\,\nu}$ with $\nu$ the correlation length exponent.
We can thus posit that $U_4$ satisfies an ansatz of the form
$
U_4$\,$=$\,$\mathcal{F}^{\,(1)}_\alpha\,(L_y^{1/\nu}\,(\delta$\,$-$\,$\delta_c)/\Omega),
$
with $\mathcal{F}$ some universal scaling function. 
Indeed, excellent data collapse \cite{landau1976finite, binder1980phase, bhattacharjee2001measure} is achieved 
upon plotting $U_4$ as a function of the scaling variable $L_y^{1/\nu}$\,$(\delta$\,$-$\,$\delta_c)$ for different values of $\delta$ and $L_y$ [Fig.~\ref{fig:Collapse}(b)],
using the exponent $\nu$\,$\approx$\,$0.629$ of the 3D (classical) Ising transition \cite{hasenbusch1999critical, el2014solving,kos2016precision}.  

Similarly, other critical exponents can be accessed from the squared staggered magnetization and susceptibility 
$
\chi^{}_s$\,$\equiv$\,$N\,(\langle M_N^2\rangle$\,$-$\,$\langle\lvert M_N^{}\rvert\rangle^2).
$
These (dimensionful) quantities obey the ans\"{a}tze \cite{troyer1997critical}
\begin{alignat}{1}
M_N^2 &=  L^{-2\beta/\nu}\mathcal{F}^{\,(2)}_\alpha \left( L_y^{1/\nu}\,(\delta-\delta_c)/\Omega \right),\\
\chi^{}_s &=  L^{2-(\eta+z)}\mathcal{F}^{\,(3)}_\alpha \left( L_y^{1/\nu}\,(\delta-\delta_c)/\Omega \right),
\end{alignat}
where $\beta$, $\eta$, and $z$ are the magnetization, anomalous spin scaling, and dynamical critical exponents, respectively, which are related as $\beta$\,$=$\,$\nu\,(\eta+D+z-2)$ \cite{cardy1996scaling}. Once again, the resulting curves for different system sizes merge into a single one using the exponents $\beta$\,$\approx$\,$0.326$,\,$\eta$\,$\approx$\,$0.036$, and $z$\,$=$\,$1$ (\textit{i.e.}, that of a Lorentz invariant theory), confirming that the QPT unequivocally belongs to the 3D Ising universality class. 

In principle, one could carry out a similar analysis for the QPTs to the other ordered phases. In practice, however, this is computationally intractable with the present DMRG approach.
%\footnote{Over our range of numerically accessible system sizes, it turns out that $\nexists$ a sequence $\{(L_x$,\,$L_y)$\,$\vert$\,$L_x/L_y$\,$=$\,$\mathrm{constant}\}$ of length $>2$ that is compatible with the magnetization profiles of all the phases, thereby rendering FSS infeasible.}. 
We therefore consider an alternative strategy, and construct effective Landau theories \cite{landau1937theory} to describe the transitions. The central idea behind this framework is similar to conventional mean-field theory, in which one represents an exponentially large number of degrees of freedom by an order parameter expectation value, say, $\psi_0$; minimizing the free energy as a function of $\psi_0$ then yields the optimal equilibrium state.
Landau theory builds upon this concept of free energy optimization by additionally incorporating spatial fluctuations, within the ``soft spin'' approximation. With reference to the Rydberg system, this means that the discrete local density $n_i$ at each site $i$ is promoted to a coarse-grained continuous magnetization field $\phi\,(\vect{r})$\,$\in$\,$\mathbb{R}$ (whose magnitude can vary freely), representing an approximate average of $n_i$ in the vicinity of $\vect{r} = \vect{r}_i$; the hard occupation constraint is thus softened. 
The underlying assumption is that the important spatial variations occur on a scale much larger than the lattice spacing \cite{sachdev2011quantum}, so we effectively focus on long-wavelength and low-energy physics.
The field $\phi\, (\vect{r})$ can be expanded in the basis set of the real-space eigenfunctions of the $\mathcal{N}$ lowest-energy modes as
\begin{alignat}{1}
\label{eq:phi}
\phi (\vect{r}) = \mathrm{Re} \left( \sum_{n=1}^{\mathcal{N}} \psi^{}_n \,\mathrm{e}^{i \vect{k}^{}_n\cdot \vect{r}}  \right),
\end{alignat}
where $\psi^{}_n$\,$\in$\,$\mathbb{C}$ is the order parameter corresponding to the $n$-th mode. The positions of these soft modes in momentum space can be identified from the peaks in the Fourier spectra listed in Table~\ref{Table:z}. The Landau functional is given by all homogeneous quartic polynomials in the amplitudes $\psi^{}_n$ that are invariant under the symmetry transformations of the underlying square lattice \cite{balents2005putting, huh2011vison, grass2011quantum, hwang2015z}. 
For example, in the case of the striated phase, our functional for the two (real) amplitudes is 
\begin{alignat}{1}
\label{eq:L1}
\mathcal{L}^{}_1 = \sum_{n=1}^2 \left(\nabla \psi_n^2 + r  \psi_n^2 \right)&+ u \sum_{n=1}^2  \psi_n ^4  + v\,  \psi_1^2\,  \psi_2^2,
\end{alignat}
where $r$,\,$u$, and $v$ are coupling constants; we need $v<0$ to ensure that both $\psi_{1,2}$ condense in the ordered phase. 
The quartic perturbation breaks the O($2$) symmetry of the quadratic terms down to $D_4$. The values of the order parameters $\psi_n$, for the different possible states, are obtained by minimizing $\mathcal{L}^{}_1$, given $r$,\,$u$, and $v$. Analyzing the RG flow of this theory provides us with valuable information about the set of fixed points and their corresponding critical exponents \cite{Pelissetto2000}, which can also be measured experimentally \cite{keesling2019quantum}. Specifically, a theory like $\mc{L}_1$ informs us of whether the transition is in the well-known $\mathscr{N}$-vector universality classes \cite{zinn1996quantum} or if the O($\mathscr{N}$)-symmetry-breaking terms modify the asymptotic critical behavior. This question of the relevance of the anisotropic perturbations can be directly addressed by classifying them \cite{Pelissetto2000, domb2001phase} using irreducible representations of the O$(\mathscr{N})$ internal group, and computing the RG dimensions of their associated couplings \cite{Vicari2011}. 

The Landau functionals for the other QPTs % to the other ordered phases 
are presented in Sec.~II of the SM and summarized in Table~SI therein \cite{sm}. These involve four or more real fields and are described by three-dimensional O($\mathscr{N}$)-symmetric vector models ($\mathscr{N}$\,$=$\,$4,\,8$) \cite{stanley1968dependence} with anisotropic perturbations \cite{Vicari2011}. While two of these Landau-Ginzburg-Wilson theories have been studied \cite{Pelissetto2000,Calabrese2004}, we also find an exotic theory % [Eq.~\eqref{eq:L4}] 
[Eq.~(S15)]
that has not been investigated previously. Experimentally extracting the exponents of the transition, possibly via the Kibble-Zurek mechanism \cite{kibble1976topology, kibble1980some, zurek1985cosmological, zurek1993cosmic, zurek1996cosmological}, can help reveal the critical properties of this theory.

Lastly, we comment on the possibility of probing the above-mentioned phase diagram experimentally. 
In an adiabatic sweep protocol (where the detuning is ramped up, driving the system from the disordered to an ordered phase), the time needed to maintain adiabaticity scales as $N/\Omega$. Due to the finite lifetime of Rydberg states, $t'$, on average, $N_0$\,$\simeq$\,$N^2/(\Omega\, t')$ atoms will spontaneously emit a photon during this process. For experiments with $N$\,$=$\,$100$ Rb atoms, Rabi frequencies of $(2\pi)$\,$\times$10\,MHz, and $t'$\,$\approx$\,$150\, \mu$s for a 70S Rydberg state, we estimate $N_0$\,$\sim$\,$1$. Such driving parameters, as well as sweeps over the required detuning range $\delta$\,$\sim$\,$2\mbox{--}3\,\Omega$ have already been utilized in one-dimensional atom arrays \cite{endres2016atom} to demonstrate physics originating from coherent many-body dynamics \cite{bernien2017probing,keesling2019quantum, levine2018high, Omran570}. 
Even though a single spontaneous emission event can affect the full many-body coherence \cite{glaetzle2017coherent,zeiher2017coherent,de2018coherent,lee2019coherent}, the small number of such expected emission events points to the promising experimental feasibility of both coherently preparing all the different density-wave-ordered ground states and observing the essential characteristics of these crystalline phases. 

 \emph{Outlook and conclusion.}---
We have numerically studied the ground-state phase diagram of interacting Rydberg atoms arranged on a 2D square lattice. We illustrated that even in this relatively simple geometry, owing to the nontrivial constraints imposed by the Rydberg blockade, a variety of intricate competing ordered phases and exotic phase transitions are realized. 
Our work serves as a useful guide to and benchmark for experiments with Rydberg atoms in 2D, and more generally, highlights the utility of Rydberg quantum simulators \cite{weimer2010rydberg} in higher dimensions as fertile test beds to explore and realize novel physical phenomena.
This begets the question: going beyond the square lattice and considering more complex geometries,
where numerical studies become ever more intractable but which are still within the grasp of a quantum simulator, what exciting new physics can we % hope to 
observe?

\begin{acknowledgments} We acknowledge useful discussions  with M.~Dalmonte, E.~Vicari, and O.~Viyuela. This research was supported by by the U.S.~Department of Energy under Grant No. $\mbox{DE-SC0019030}$, the National Science Foundation (NSF), the Harvard--MIT Center for Ultracold Atoms, the Office of Naval Research, and the Vannevar Bush Faculty Fellowship. The computations in this paper were run on the FASRC Cannon and Odyssey clusters supported by the FAS Division of Science Research Computing Group at Harvard University. R.S.~thanks J.~Coulter and E.M.~Stoudenmire for computational support. W.W.H.~is supported by the Gordon and Betty Moore Foundation's EPiQS Initiative, Grant No.~GBMF4306, and the NUS Development Grant AY2019/2020. H.P.~was supported by the NSF through a grant for the Institute for Theoretical Atomic, Molecular and Optical Physics at Harvard University and the Smithsonian Astrophysical Observatory, and by the Gordon and Betty Moore Foundation's EPiQS Initiative, Grant No.~GBMF8682.
\end{acknowledgments}

\bibliographystyle{apsrev4-1}
\bibliography{Ryd2D_Refs}
%\bibliography{Draft4.bbl}



\section*{Supplementary material}

# Supplemental Material for "Complex density wave orders and quantum phase transitions in a model of square-lattice Rydberg atom arrays"

Rhine Samajdar,[1] Wen Wei Ho,[1] Hannes Pichler,[1,2,3] Mikhail D. Lukin,[1] and Subir Sachdev[1]

[1]*Department of Physics, Harvard University, Cambridge, MA 02138, USA*
[2]*ITAMP, Harvard-Smithsonian Center for Astrophysics, Cambridge, MA 02138, USA*
[3]*Division of Physics, Mathematics and Astronomy,
California Institute of Technology, Pasadena, CA 91125, USA*

In this Supplemental Material, we provide details of the computational methods used for the calculations described in the main text. We also derive and discuss the effective Landau theories for the quantum phase transitions in the 2D Rydberg atom array.

## I. COMPUTATIONAL METHODS

The primary tool that we employ to find the ground states of $H_{\mathrm{Ryd}}$ is the finite-system density-matrix renormalization group (DMRG) [1–4], implemented using the ITensor package [5]. The desired wavefunction can be represented as a matrix product state (MPS) [6, 7] of the form

$$|\Psi\rangle = \sum_{\tau_1\ldots\tau_n} \sum_{b_1\ldots b_{n-1}} A^{\tau_1}_{b_1} A^{\tau_2}_{b_1 b_2} A^{\tau_3}_{b_2 b_3} \cdots A^{\tau_n}_{b_{n-1}} |\tau_1,\ldots,\tau_n\rangle, \tag{S1}$$

where the $A$ are matrices with physical indices $\tau$ and link indices $b$. The DMRG algorithm then presents an efficient method to find the optimal MPS representation of the many-body state [8].

Unless stated otherwise, we place the system on a cylinder with open boundary conditions along $\hat{x}$ but periodic along $\hat{y}$, as opposed to fully periodic boundaries, which would necessitate squaring the number of states required for a given accuracy [9]. One of the major drawbacks of two-dimensional DMRG is that the number of states retained must be increased exponentially with the width of the system to maintain a constant accuracy [10] and this constrains the system sizes that can be simulated: in this work, we considered sizes ranging from $L_x = 8$–$20$ and $L_y = 4$–$10$ with an associated bond dimension $d = 200$–$1600$. The system is regarded to have converged to its true ground state once the truncation error drops below a certain threshold value (taken to be $10^{-11}$ here) and in practice, the convergence criterion was typically found to be satisfied after performing $\sim \mathcal{O}(10^2)$ successive sweeps. Our sweeping strategy entails initially performing a large number of sweeps at relatively small bond dimensions before ramping $d$ up progressively at later stages. To help facilitate the proper build-up of long-range correlations, a small "noise" term [11] is initially added to the density matrix at each step, but then turned off in later sweeps. Given a ground state $|\psi_0\rangle$ obtained in this fashion, we can also target the first-excited state using the Hamiltonian $H' = H_{\mathrm{Ryd}} + wP_0$, where $P_0 = |\psi_0\rangle\langle\psi_0|$ is a projection operator and $w$ is an energy penalty.

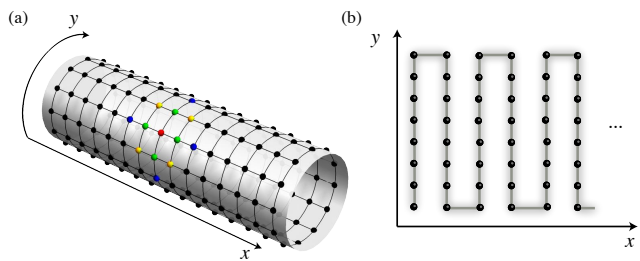

FIG. S1. (a) The real-space geometry of the system used for performing the DMRG computations, illustrating the cylindrical boundary conditions. The $x$-axis is the direction along the cylinder, while the $y$-axis wraps around it. (b) The numbering scheme used to transform the 2D lattice into a 1D chain results in a so-called snake-like ordering.

Particular care must be taken to ensure the compatibility of the density-wave ordering with the system size and boundary conditions. Since the open boundaries act as effective pinning fields [12, 13], the ground state can differ nontrivially between lattices with even and odd lengths. For instance, both the striated and star phases require $L_x$ to be odd—because the system prefers maximizing Rydberg excitations at the edges—but on such lattices, we cannot isolate a defect-free state belonging to the rhombic phase due to the very same reason. A signature of this disagreement with our $15 \times 8$ system size is seen in the small fluctuations of the entanglement entropy in the top-right corner of the phase diagram in Fig. 1 of the main text. In the same vein, as evidenced by Fig. 3(a), the ideal local magnetization pattern of the staggered phase is not compatible with the combination of cylindrical boundaries and even $L_y$, so the numerically calculated ground state will always have a nonzero density of defects. To reduce these boundary effects, one can either study the central bulk of a given $L_x \times L_y$ system [14, 15] or switch to open boundaries.

## II. DERIVATION OF THE LANDAU FUNCTIONALS

In the spirit of Refs. [16–19], in this section, we formulate the most general continuum theory of the order-parameter fields $\psi_n$ that is invariant under the symmetry transformations of the square lattice. The elements of the space group include translations along the $x$ ($T_x$) and $y$ ($T_y$) axes, reflections about the $x$ ($R_x$) and $y$ ($R_y$) axes as well as the two diagonals $x = y$ ($R_d$) and $x = -y$ ($R_{d'}$), and fourfold rotations around the $z$ axis ($C_4$). To write down the quartic Landau functional, we need to determine how the $\psi_n$ transform under these operations. This, in turn, follows from the transformation properties of the eigenvectors $\exp(i\boldsymbol{k}_n \cdot \boldsymbol{r})$, introduced in Eq. (4) of the main text, as

$$\hat{\mathcal{O}}\phi(\boldsymbol{r}) = \operatorname{Re}\left[\sum_n \psi_n e^{i\boldsymbol{k}_n \cdot (\hat{\mathcal{O}}\boldsymbol{r})}\right] \equiv \operatorname{Re}\left[\sum_n \left(\hat{\mathcal{O}}\psi\right)_n e^{i\boldsymbol{k}_n \cdot \boldsymbol{r}}\right].$$

We outline this procedure individually for each of the phases in the following.

### A. Disordered to striated

The minimal set of momenta $\{\boldsymbol{k}_n\}$ required to describe the magnetization field $\phi(\boldsymbol{r})$ can be read off Table I of the main text. In the striated phase, $|n(\boldsymbol{k})|$ shows peaks at not only $(\pi, 0)$ and $(0, \pi)$ but also $(\pi, \pi) = (\pi, 0) + (0, \pi)$. For the purpose of describing the phase transition, therefore, it suffices to focus on the first two momenta alone. The magnetization can be expressed in terms of two *real* fields $\psi_1$ and $\psi_2$ as

$$\phi(\boldsymbol{r}) = \psi_1 e^{i(\pi, 0) \cdot \boldsymbol{r}} + \psi_2 e^{i(0, \pi) \cdot \boldsymbol{r}}. \quad (S2)$$

In the basis $\Psi \equiv (\psi_1, \psi_2)$, the matrix representations of the symmetry transformations are:

$$T_x = -\sigma_3,\ T_y = \sigma_3,\ R_x = R_y = \mathbb{1},\ R_d = R_{d'} = C_4 = \sigma_1,$$

where $\sigma$ are the $2 \times 2$ Pauli matrices. The Landau functional is given by all homogeneous polynomials that are invariant under the group generated by these transformations and, up to quartic order, is expressed by Eq. (5) of the main text.

### B. Disordered to star

The Fourier maxima in this phase are seen to be at $(\pi, 0)$, $(0, \pi)$, $(\pi/2, \pi)$, and $(\pi, \pi/2)$, but noting that $(\pi, 0) = 2(\pi/2, \pi)$—and similarly for $(0, \pi)$—we can write the magnetization as simply

$$\phi(\boldsymbol{r}) = \operatorname{Re}\left(\psi_1 e^{i(\pi/2, \pi) \cdot \boldsymbol{r}} + \psi_2 e^{i(\pi, \pi/2) \cdot \boldsymbol{r}}\right), \quad (S3)$$

and the other wavevectors are described by harmonics, $\psi_{1,2}^2$, of the order parameters. Unlike in the previous case, $\psi_1$ and $\psi_2$ are now complex. Using the basis, $\Psi \equiv (\psi_1, \psi_2, \psi_1^*, \psi_2^*)$, the symmetry operators can be written as matrices which are

$$T_x = \begin{bmatrix} i & 0 & 0 & 0 \\ 0 & -1 & 0 & 0 \\ 0 & 0 & -i & 0 \\ 0 & 0 & 0 & -1 \end{bmatrix},\ T_y = \begin{bmatrix} -1 & 0 & 0 & 0 \\ 0 & i & 0 & 0 \\ 0 & 0 & -1 & 0 \\ 0 & 0 & 0 & -i \end{bmatrix}, \quad (S4)$$

$$R_x = \begin{bmatrix} 1 & 0 & 0 & 0 \\ 0 & 0 & 0 & 1 \\ 0 & 0 & 1 & 0 \\ 0 & 1 & 0 & 0 \end{bmatrix},\ R_y = \begin{bmatrix} 0 & 0 & 1 & 0 \\ 0 & 1 & 0 & 0 \\ 1 & 0 & 0 & 0 \\ 0 & 0 & 0 & 1 \end{bmatrix},$$

$$R_d = \begin{bmatrix} 0 & 1 & 0 & 0 \\ 1 & 0 & 0 & 0 \\ 0 & 0 & 0 & 1 \\ 0 & 0 & 1 & 0 \end{bmatrix},\ R_{d'} = \begin{bmatrix} 0 & 0 & 0 & 1 \\ 0 & 0 & 1 & 0 \\ 0 & 1 & 0 & 0 \\ 1 & 0 & 0 & 0 \end{bmatrix},\ C_4 = \begin{bmatrix} 0 & 0 & 0 & 1 \\ 1 & 0 & 0 & 0 \\ 0 & 1 & 0 & 0 \\ 0 & 0 & 1 & 0 \end{bmatrix}.$$

These seven matrices generate a subgroup of O(4) and the quartic terms of the Landau functional, composed of all polynomials invariant under this group, is written as

$$\bar{\mathcal{L}}_2 = \sum_{n=1}^{2}\left[u\,|\psi_n|^4 + w\left(\psi_n^4 + (\psi_n^*)^4\right)\right] + v\,|\psi_1|^2|\psi_2|^2. \quad (S5)$$

We need $v > 0$ to ensure only one of $\psi_{1,2}$ condenses, as observed in the star phase. This model is equivalent to the so-called tetragonal theory which is the $M = 2$, $N = 2$ version of the general three-coupling Landau-Ginzburg-Wilson Hamiltonian

$$\mathcal{H} = \int d^dx \Big\{ \sum_{i,a} \frac{1}{2}\left[(\partial_\mu \phi_{a,i})^2 + r\phi_{a,i}^2\right] \quad (S6)$$
$$+ \sum_{ij,ab} \frac{1}{4!} \left(u_0 + v_0 \delta_{ij} + w_0 \delta_{ij}\delta_{ab}\right) \phi_{a,i}^2 \phi_{b,j}^2 \Big\},$$

where $a, b = 1, ...M$ and $i, j = 1, ...N$. The RG flow of this model was discussed in Sec. 11.6 of Ref. [20]. Besides the eight fixed points [21–23] referenced therein, another important fixed point—relevant to the case at hand—is given by the chiral O(2)×O(2) theory [24], which is obtained upon setting one of the couplings to zero. It is uncertain whether this chiral fixed point is stable in the enlarged tetragonal theory.

### C. Disordered to banded

The transition to the banded phase involves the onset of Fourier peaks at $\pm(2\pi/5, \pi)$, which continue to persist in the rhombic phase at larger $\delta/\Omega$. Accordingly, we write

$$\phi(\boldsymbol{r}) = \operatorname{Re}\left(\psi_1 e^{i(2\pi/5, \pi) \cdot \boldsymbol{r}} + \psi_2 e^{i(\pi, 2\pi/5) \cdot \boldsymbol{r}}\right). \quad (S7)$$

In the basis, $\Psi \equiv (\psi_1, \psi_2, \psi_1^*, \psi_2^*)$, the representations of the symmetries are

$$T_x = \begin{bmatrix} \zeta & 0 & 0 & 0 \\ 0 & -1 & 0 & 0 \\ 0 & 0 & \zeta^4 & 0 \\ 0 & 0 & 0 & -1 \end{bmatrix}, T_y = \begin{bmatrix} -1 & 0 & 0 & 0 \\ 0 & \zeta & 0 & 0 \\ 0 & 0 & -1 & 0 \\ 0 & 0 & 0 & \zeta^4 \end{bmatrix}, \quad \text{(S8)}$$

$$R_x = \begin{bmatrix} 1 & 0 & 0 & 0 \\ 0 & 0 & 0 & 1 \\ 0 & 0 & 1 & 0 \\ 0 & 1 & 0 & 0 \end{bmatrix}, R_y = \begin{bmatrix} 0 & 0 & 1 & 0 \\ 0 & 1 & 0 & 0 \\ 1 & 0 & 0 & 0 \\ 0 & 0 & 0 & 1 \end{bmatrix},$$

$$R_d = \begin{bmatrix} 0 & 1 & 0 & 0 \\ 1 & 0 & 0 & 0 \\ 0 & 0 & 0 & 1 \\ 0 & 0 & 1 & 0 \end{bmatrix}, R_{d'} = \begin{bmatrix} 0 & 0 & 0 & 1 \\ 0 & 0 & 1 & 0 \\ 0 & 1 & 0 & 0 \\ 1 & 0 & 0 & 0 \end{bmatrix}, C_4 = \begin{bmatrix} 0 & 0 & 0 & 1 \\ 1 & 0 & 0 & 0 \\ 0 & 1 & 0 & 0 \\ 0 & 0 & 1 & 0 \end{bmatrix},$$

employing the shorthand $\zeta = \exp(2\pi i/5)$. Repeating the same procedure as above leads to the Landau functional

$$\bar{\mathcal{L}}_3 = \sum_{n=1}^{2} u|\psi_n|^4 + v|\psi_1|^2|\psi_2|^2, \quad \text{(S9)}$$

where we have suppressed the trivial quadratic and derivative terms. Once again, $v > 0$ is needed to ensure that only one of $\psi_{1,2}$ condenses in the banded state. This is a particular case of the $MN$ model [25, 26] (the Hamiltonian for which is given by Eq. (S6) for $w_0 = 0$), with $M = 2$ and $N = 2$, or equivalently, of the O(2) × O(2) chiral model [27, 28],

$$\mathcal{H} = \int d^d x \Big\{ \frac{1}{2} \sum_a \Big[ (\partial_\mu \phi_a)^2 + r\phi_a^2 \Big] + \frac{1}{4!} u_0 \Big( \sum_a \phi_a^2 \Big)^2$$
$$+ \frac{1}{4!} v_0 \sum_{a,b} \Big[ (\phi_a \cdot \phi_b)^2 - \phi_a^2 \phi_b^2 \Big] \Big\}, \quad \text{(S10)}$$

which is related to the former by an exact mapping of the fields. It has two stable fixed point in different regions of the quartic parameters, divided by a separatrix, and the RG flow leading to the "chiral" fixed point was studied in Ref. [29]. One can show that this theory also has a XY fixed point in another region of the quartic couplings.

### D. Banded to rhombic

Going into the rhombic phase, $|n(\boldsymbol{k})|$ develops additional maxima at $\pm(\pi, \pi/4)$. Focusing on these dominant peaks, we formulate the field as

$$\phi(\boldsymbol{r}) = \text{Re}\left( \psi_1 e^{i(\pi/4, \pi) \cdot \boldsymbol{r}} + \psi_2 e^{i(\pi, \pi/4) \cdot \boldsymbol{r}} \right). \quad \text{(S11)}$$

As before, the matrix representations of the symmetries, in the basis, $\Psi \equiv (\psi_1, \psi_2, \psi_1^*, \psi_2^*)$, are

$$T_x = \begin{bmatrix} \lambda & 0 & 0 & 0 \\ 0 & -1 & 0 & 0 \\ 0 & 0 & \lambda^* & 0 \\ 0 & 0 & 0 & -1 \end{bmatrix}, T_y = \begin{bmatrix} -1 & 0 & 0 & 0 \\ 0 & \lambda & 0 & 0 \\ 0 & 0 & -1 & 0 \\ 0 & 0 & 0 & \lambda^* \end{bmatrix}, \quad \text{(S12)}$$

$$R_x = \begin{bmatrix} 1 & 0 & 0 & 0 \\ 0 & 0 & 0 & 1 \\ 0 & 0 & 1 & 0 \\ 0 & 1 & 0 & 0 \end{bmatrix}, R_y = \begin{bmatrix} 0 & 0 & 1 & 0 \\ 0 & 1 & 0 & 0 \\ 1 & 0 & 0 & 0 \\ 0 & 0 & 0 & 1 \end{bmatrix},$$

$$R_d = \begin{bmatrix} 0 & 1 & 0 & 0 \\ 1 & 0 & 0 & 0 \\ 0 & 0 & 0 & 1 \\ 0 & 0 & 1 & 0 \end{bmatrix}, R_{d'} = \begin{bmatrix} 0 & 0 & 0 & 1 \\ 0 & 0 & 1 & 0 \\ 0 & 1 & 0 & 0 \\ 1 & 0 & 0 & 0 \end{bmatrix}, C_4 = \begin{bmatrix} 0 & 0 & 0 & 1 \\ 1 & 0 & 0 & 0 \\ 0 & 1 & 0 & 0 \\ 0 & 0 & 1 & 0 \end{bmatrix},$$

where $\lambda$ denotes $\exp(i\pi/4)$. The final form of the quartic polynomial obtained by imposing invariance under these symmetries is exactly the same as Eq. (S9).

### E. Disordered to staggered

Table. I of the main text arrays a set of four independent Fourier peaks for the staggered phase in terms of which, the magnetization field is

$$\phi(\boldsymbol{r}) = \text{Re}\bigg( \psi_1 e^{i(2\pi/5, 6\pi/5) \cdot \boldsymbol{r}} + \psi_2 e^{i(4\pi/5, 2\pi/5) \cdot \boldsymbol{r}}$$
$$+ \psi_3 e^{i(6\pi/5, 2\pi/5) \cdot \boldsymbol{r}} + \psi_4 e^{i(2\pi/5, 4\pi/5) \cdot \boldsymbol{r}} \bigg). \quad \text{(S13)}$$

For notational brevity, we refrain from expressing the symmetry transformations as explicit $8 \times 8$ matrices, and instead just list their operations on the $\psi_n$:

$$\begin{aligned}
T_x &: \psi_1 \to \zeta\psi_1, \ \psi_2 \to \zeta^2\psi_2, \psi_3 \to \zeta^3\psi_3, \psi_4 \to \zeta\psi_4; \\
T_y &: \psi_1 \to \zeta^3\psi_1, \psi_2 \to \zeta\psi_2, \ \psi_3 \to \zeta\psi_3, \ \psi_4 \to \zeta^2\psi_4; \\
R_x &: \psi_1 \to \psi_4, \quad \psi_2 \to \psi_3^*, \quad \psi_3 \to \psi_2^*, \quad \psi_4 \to \psi_1; \\
R_y &: \psi_1 \to \psi_4^*, \quad \psi_2 \to \psi_3, \quad \psi_3 \to \psi_2, \quad \psi_4 \to \psi_1^*; \\
R_d &: \psi_1 \to \psi_3, \quad \psi_2 \to \psi_4, \quad \psi_3 \to \psi_1, \quad \psi_4 \to \psi_2; \\
R_{d'} &: \psi_1 \to \psi_3^*, \quad \psi_2 \to \psi_4^*, \quad \psi_3 \to \psi_1^*, \quad \psi_4 \to \psi_2^*; \\
C_4 &: \psi_1 \to \psi_2^*, \quad \psi_2 \to \psi_1, \quad \psi_3 \to \psi_4, \quad \psi_4 \to \psi_3^*.
\end{aligned} \quad \text{(S14)}$$

The most general quartic polynomial consistent with these transformations is

$$\bar{\mathcal{L}}_4 = \sum_{n=1}^{4} u|\psi_n|^4 + v_1 \left( |\psi_1|^2|\psi_2|^2 + |\psi_3|^2|\psi_4|^2 \right) \quad \text{(S15)}$$
$$+ v_2 \left( |\psi_1|^2|\psi_3|^2 + |\psi_2|^2|\psi_4|^2 \right) + v_3 \left( |\psi_1|^2|\psi_4|^2 + |\psi_2|^2|\psi_3|^2 \right)$$
$$+ w \left[ \psi_1 (\psi_2^*)^3 + \psi_1^3 \psi_2 + \psi_3^3 \psi_4 + \psi_3 (\psi_4^*)^3 + \text{c.c} \right].$$

While the critical exponents of $N$-vector models have been estimated for large $N$ using $1/N$ expansions and

| Phase | Landau theory | Fixed points |
|---|---|---|
| Striated | 2-vector model with $D_4$ symmetry | Ref. [33] |
| Star | Tetragonal theory [20] | Refs. [21–24] |
| Banded | $MN$ model ($M=N=2$) [25, 26]/ $O(2) \times O(2)$ chiral model [27, 28] | Ref. [29] |
| Staggered | 3D O(8) model + anisotropic perturbations | — |

TABLE SI. Summary of the Landau theories for the QPTs from the disordered phase to the various ordered phases, and the studies that describe their fixed points. Note that the transition from the banded to the rhombic phase is described by the same theory as that from the disordered to the banded, and is, therefore, omitted above.

RG techniques [30–32], the quartic terms listed above break the O(8) symmetry in the present case. The resultant model of Eq. (S15) has not been analyzed previously in the literature and its critical behavior remains an open question, which we leave for future work.

Finally, we close by recapitulating our results of this section in Table SI for convenience of reference.



\end{document}